\documentclass[apj]{emulateapj}
\usepackage{graphicx}
\usepackage{epstopdf}
\usepackage{amsmath}
\usepackage{amssymb}
\shorttitle{Neutron star accretion flow}
\shortauthors{Bu, Qiao, \& Yang}
\begin{document}
\title{Two-temperature radiative hot accretion flow around neutron stars}
\author{De-Fu Bu\altaffilmark{1}, Erlin Qiao\altaffilmark{2,3} Xiao-Hong Yang\altaffilmark{4}}

\altaffiltext{1}{Corresponding author\\Key Laboratory for Research in Galaxies and Cosmology, Shanghai Astronomical Observatory, Chinese Academy of Sciences, 80 Nandan Road, Shanghai 200030, China; dfbu@shao.ac.cn }
\altaffiltext{2}{Corresponding author\\Key Laboratory of Space Astronomy and Technology, National Astronomical Observatory, Chinese Academy of Sciences, Beijing 100012, China; qiaoel@nao.cas.cn }
\altaffiltext{3}{School of Astronomy and Space Sciences, University of Chinese Academy of Sciences, 19A Yuquan Road, Beijing 100049 }
\altaffiltext{4}{Department of physics, Chongqing University, Chongqing, 400044}

%\date{Accepted 1988 December 15. Received 1988 December 14; in original form 1988 October 11}

%\pagerange{\pageref{firstpage}--\pageref{lastpage}} \pubyear{2002}

%\maketitle

%\label{firstpage}
%
\begin{abstract}
Numerical simulations of radiative two-temperature hot accretion flows (HAFs) around Neutron stars (NSs) are performed. We assume that all of the energy carried by the HAF around a NS will be thermalized and radiated out at the surface of the NS. The thermal photons will propagate outwards radially and cool the HAF vis Comptonization. We define $\dot m$ as the mass accretion rate at the surface of the central object in unit of Eddington accretion rate ($\dot M_{\rm Edd}=10L_{\rm Edd}/c^2$, with $L_{\rm Edd}$ and $c$ being Eddington luminosity and speed of light, respectively). When $\dot m$ is lower than $\sim 10^{-4}$, the cooling of the HAF is not important and outflows are very strong. When $\dot m > \sim 10^{-3}$, cooling becomes important and outflows are significantly weak. In the range $10^{-4} < \dot m < 10^{-3}$, the HAFs transients from a strong outflow phase to a very weak outflow phase with increase of $\dot m$. The properties of the HAF around a NS are also compared to those of the HAF around a BH. We find that with a similar $\dot m$, the dynamical properties of the HAF around a NS are quite similar as those of the HAF around a BH. However, the emitted spectrum of a HAF around a NS can be quite different from that of a HAF around a BH due to the presence of a thermal soft X-ray component coming from the surface of the NS.
\end{abstract}


\section{Introduction}
In a X-ray binary system, a compact object accretes gas from its companion star. In the black hole low mass X-ray binaries (BH-LMXBs) and the neutron star low mass X-ray binaries (NS-LMXBs), the companion star supplies gas to the compact object through the Lagrangian point. The accretion flow model for the low/hard spectral state of both BH-LMXBs and NS-LMXBs is a hot accretion flow (see Yuan \& Narayan 2014 for reviews). The accretion disk model for the high/soft spectral state of both BH-LMXBs and NS-LMXBs is a standard thin disk (Shakura \& Sunyaev 1973). The transition from the low/hard spectral state to the high/soft spectral state occurs when the luminosity of the system crosses a critical luminosity. The value of the critical luminosity is $\sim (1 - 4)\% L_{\rm Edd}$ (e.g., Nowak et al. 2002; Maccarone \& Coppi 2003; Rodriguez et al. 2003; Kubota \& Done 2004;  Meyer-Hofmeister et al. 2005; Gladstone et al. 2007; Qiao \& Liu 2009; Qiao \& Liu 2013; Zhang et al. 2016).

The inner boundary conditions of a HAF around a NS are different from those of a HAF around a BH. For a BH accretion flow, all the energy carried by the accretion flow will go into the event horizon. A NS has a hard surface. Therefore, one can expect that all the energy carried by the accretion flow around a NS will be radiated out at the surface of the NS (Zampieri et al. 1995).  The radiation at the surface of a NS will propagate outwards through the accretion flow. Consequently, the accretion flow will be cooled down by inverse Compton scattering. Therefore, at a same accretion rate, the electron temperature of a HAF around a NS will be lower than that of a HAF around a BH. This point has been confirmed by analytical (Sunyaev \& Titarchuk 1989; Syunyaev et al. 1991; Narayan \& McClintock 2008; Qiao \& Liu 2018) and simulation (Bu, Qiao \& Yang 2019) works. The observations to BH-LMXBs and NS-LMXBs do show that electron temperature of a HAF around a NS is lower than that of a HAF around a BH (Burke et al. 2017). There are some other differences between the properties of accretion flows around a NS and a BH. First, observations to NS-LMXBs show that in the luminosity range $(0.01\% - 1\%) L_{\rm Edd}$, there is a thermal soft X-ray component (e.g., Jonker et al. 2004; Armas Padilla et al. 2013a, 2013b; Degenaar et al. 2013b; Campana et al. 2014). The thermal soft X-ray component is produced at the surface of the NS (Degenaar et al. 2013a; Hernandez Santisteban et al. 2018; van den Eijnden et al. 2018). In the similar luminosity range, a thermal soft X-ray component has not been observed for BH-LMXBs (Degenaar et al. 2013b; Bahramian et al. 2014). Second, due to the presence of the thermal soft X-ray component in the NS-LMXB systems, it is expected that at a same accretion rate, a NS-LMXB system should be brighter than a BH-LMXB system. This point has been confirmed by observations (e.g., Menou et al. 1999; Lasota 2000; Garcia et al. 2001; Hameury et al. 2003; McClintock et al. 2004).

There are many simulation works studying the properties of HAFs around BHs (e.g., Stone et al. 1999; Igumenshchev \& Abramowicz 2000; Tchekhovskoy et al. 2011; Narayan et al. 2012; Yuan et al. 2012, 2015; Bu et al. 2016a, 2016b; Bu \& Gan 2018). However, there are very few simulation works studying the properties of HAFs around NSs. Very recently, Bu, Qiao, \& Yang (2019, hereafter Paper I) performed simulations of HAFs around NSs. In that work, we assume that electrons and ions have same temperatures. Also, in that work, the radiative cooling processes of the HAF (e.g., bremsstrahlung, synchrotron and their self-Comptonizations) are neglected. In Paper I, we only focus on the effects of inver-Compton cooling due to the presence of themal photons from the surface of a NS on the properties of a HAF. It is found that when $\dot m>10^{-4}$, the HAF can be cooled efficiently by the inverse-Compton scattering between gas and photons from the surface of the NS. Consequently, outflows/winds disappear due to very weak gas pressure gradient force.

In this paper, we perform two-temperature radiative simulations of HAFs around NSs. The new points in the present paper are as follows. First, the ions and electrons have different temperatures. The ions can give energy to electrons by Coulomb collision. Second, electrons can cool by radiative coolings (bremsstrahlung, synchrotron and the self-Comptonizations). Based on the properties of the HAFs around NSs obtained in this paper, one can calculate the spectrum of a NS-LMXB system and compare it to observations.

We organize the paper as follows. The numerical settings are introduced in Section 2. In Section 3, we present our results. We summarize our results in Section 4.

\section{Numerical method }
By using the ZEUS-3D code (Clarke 1996), we perform two-dimensional hydrodynamic simulations of two-temperature radiative HAFs around both NSs and BHs. The black hole mass is set to be $M_{\rm BH}=10M_{\odot}$, with $M_{\odot}$ being the solar mass. The mass of the neutron star is $M_{\rm NS}=1.4 M_{\odot}$.
Spherical coordinates ($r,\theta,\phi$) are employed in this paper. The equations describing the accretion flow are as follows,

\begin{equation}
 \frac{d\rho}{dt} + \rho \nabla \cdot {\bf v} = 0,
\end{equation}
\begin{equation}
 \rho \frac{d{\bf v}}{dt} = -\nabla p_{\rm tot} - \rho \nabla \Phi + \nabla \cdot {\bf T}
\end{equation}
\begin{equation}
 \rho \frac{d(e_{\rm i}/\rho)}{dt} = -p_{\rm i}\nabla \cdot {\bf v} + (1-\delta) {\bf T}^2/\mu - q^{\rm ie}
\end{equation}
\begin{equation}
 \rho \frac{d(e_{\rm e}/\rho)}{dt} = -p_{\rm e}\nabla \cdot {\bf v} + \delta {\bf T}^2/\mu + q^{\rm ie} - Q^- + Sc
\end{equation}

$\rho$ and $\bf v$ are density and velocity. $p_{\rm tot}$ is the total pressure, $p_{\rm tot}=p_{\rm gas}+ p_{\rm mag}$, with $p_{\rm gas}$ and $p_{\rm mag}$ being gas and magnetic pressure, respectively. $e_{\rm i}$ and $e_{\rm e}$ are the internal energies of ions and electrons, respectively. $p_{\rm i}$ and $p_{\rm e}$ are pressures of ions and electrons, respectively. The total gas pressure is $p_{\rm gas}=p_{\rm i}+p_{\rm e}$. We adopt ideal gas equations of state $p_{\rm i}=(\gamma_{\rm i}-1)e_{\rm i}$ and $p_{\rm e}=(\gamma_{\rm e}-1)e_{\rm e}$. We set $\gamma_{\rm i}=5/3$ and $\gamma_{\rm e}=4/3$. We set $p_{\rm mag}=\beta p_{\rm tot} $. For NS-LMXBs, observations show that magnetic field is weak (Vaughan et al. 1994; Maccarone \& Coppi 2003; Done et al. 2007). Therefore, in this paper, we set a weak magnetic field with $\beta=0.05$.

For the simulations of HAFs around a NS, we set the gravitational potential $\Phi=GM_{\rm NS}/(r-r_s)$, with $G$ being the gravitational constant, and $r_s=2GM_{\rm NS}/c^2$ being Schwarzschild radius. For the simulations of HAFs around a BH, we set $\Phi=GM_{\rm BH}/(r-r_s)$, with $r_s=2GM_{\rm BH}/c^2$ being Schwarzschild radius.

The last term ($Sc$) in Equation (4) is the inverse Compton cooling of the accretion flow by soft photons from the surface of a NS.

In accretion flows, it is generally believed that angular momentum is transferred by Maxwell stress. When weak and tangled magnetic field is present, angular momentum is transferred by magnetohydrodynamic turbulence induced by magneto-rotational instability (e.g., Balbus \& Hawley 1991). The Maxwell stress exerted by the weak tangled magnetic field can be well described as a viscous stress tensor. It has been shown by observations that generally NS-LMXBs are just weakly magnetized (with magnetic field $<10^8$ Gauss; Vaughan et al. 1994; Maccarone \& Coppi 2003; Done et al. 2007). Thus, we can use a viscous stress tensor to mimic the angular momentum transfer by weak and tangled magnetic field.
In Equations (2), (3) and (4), $\bf T$ is the viscous stress tensor. Following Stone et al. (1999), we assume that the viscous tensor only has azimuthal components:
\begin{equation}
T_{r\phi} = \mu r \frac{\partial}{\partial r} \left( \frac{v_\phi}{r} \right)
\end{equation}
\begin{equation}
T_{\theta\phi} = \frac{\mu \sin \theta}{r} \frac{\partial}{\partial \theta} \left( \frac{v_\phi}{\sin \theta} \right)
\end{equation}
In the equations, $\mu=\rho \nu$. Following Stone et al. (1999), we assume that $\nu=\alpha \sqrt{GM} r^{1/2}$, which is the usual ``$\alpha$" description and set $\alpha=0.02$.

The parameter $\delta$ in Equations (3) and (4) denotes the fraction of the viscous heating energy that directly goes into electrons. In this paper, we set $\delta=0$, all the viscous heating energy goes directly into ions.

\subsection{Radiative cooling}
The ions and electrons interchange energy through Coulomb collision ($q^{\rm ie}$). The formula for $q^{\rm ie}$ is given by Narayan \& Yi (1995, hereafter NY95). The electrons can cool by bremsstrahlung radiation ($q_{\rm brem}^{-}$), synchrotron radiation ($q_{\rm syn}^{-}$), the self-Comptonization of bremsstrahlung radiation ($q_{\rm brem, C}^{-}$), and the self-Comptonization of synchrotron radiation ($q_{\rm syn, C}^{-}$). Therefore, the electron cooling term in Equation (4) can be written as $Q^-=q_{\rm brem}^{-}+q_{\rm syn}^{-}+q_{\rm brem, C}^{-}+q_{\rm syn, C}^{-}$. The expressions for ($q_{\rm brem}^{-}$), ($q_{\rm syn}^{-}$), ($q_{\rm brem, C}^{-}$) and ($q_{\rm syn, C}^{-}$) are given in NY95. As in Bu \& Gan (2018), we use a Compton enhancement factor $\eta$ to calculate the terms $q_{\rm brem, C}^{-}$ and $q_{\rm syn, C}^{-}$. Note that the optical depth needed for calculating $\eta$ at any radii is calculated from $\theta=0$ to $\pi$.

\begin{table*} \caption{Simulation parameters and results }
\setlength{\tabcolsep}{4mm}{
\begin{tabular}{ccccc}
\hline \hline
 Models & Central object & $\rho_{\rm max}$   & Compton temperature $T_*$ &  Mass inflow rate at  \\

  &   &   ($10^{-8}\text{g cm}^{-3}$) & &   inner boundary ($\dot M_{\rm in} (r_{\rm in})/\dot M_{\rm Edd}$)       \\
(1) & (2)             & (3)                       &     (4)  & (5)           \\

\hline\noalign{\smallskip}
NS1   & Neutron star & 3 & $7.1 \times 10^6$ K & $1.1\times 10^{-2}$    \\
NS2   & Neutron star & 1   & $5.3 \times 10^6$ K & $3 \times 10^{-3}$   \\
NS3   & Neutron star & 0.3   & $2.6\times 10^6 $ K & $1.6\times 10^{-4}$   \\
BH1   & Black hole & 26  & & $1 \times 10^{-2}$   \\
BH2   & Black hole & 13  & & $3.7 \times 10^{-3}$   \\
BH3   & Black hole & 2  &  & $1.7\times 10^{-4}$   \\

\hline\noalign{\smallskip}
\end{tabular}}

Note: Col. 1: model names. Col. 3: The maximum density at the initial torus center in unit of $10^{-8}\text{g cm}^{-3}$. Col. 4: Compton temperature of photons emitted at the surface of the NS (Equation (\ref{Tc})).  Col. 5: time-averaged mass accretion rate of the central object (in unit of Eddington rate).

\end{table*}

There are complicated interactions between the hard surface of a NS and the accretion flow. The interactions are hot topics and quite complicated. For example, when an accretion flow approaches the surface of a NS, the rotational velocity of the accretion flow should be decreased to the value equal to that of the NS. How to decrease the rotational velocity of the accretion flow is a hot topic (Medvedev \& Narayan 2001; Medvedev 2004; Gilfanov \& Sunyaev 2014; Popham \& Narayan 1992; Narayan \& Popham 1993; Inogamov \& Sunyaev 1999; Popham \& Sunyaev 2001; Burke et al. 2018). In the present paper, we simplify the interactions between a HAF and the surface of a NS. We just assume that when a HAF arrives at the surface of a NS, all of the energy carried by the HAF will be thermalized as blackbody emission and radiated out spherically (see Qiao \& Liu 2018; Paper I). The thermal photons go outwards through the HAF and cool the HAF via Comptonization.
The energy flux carried by the accretion flow onto the surface of a NS can be expressed as,
\begin{equation}
L_*=2 \pi R_*^2 \int_0^\pi \rho_{*} \min [v_{r*},0] \left[ \frac{1}{2}v^2_{*}+e_*/\rho_{*} \right] \sin\theta d\theta
\label{luminosity}
\end{equation}
where, $\rho_{*}$, $v_{r*}$, $v_{*}$ and $e_{*}$ are density, radial velocity, total velocity, and internal energy at the surface of a NS ($R_*$), respectively. The radius for a NS with $M_{\rm NS}=1.4M_\odot$ is $\sim 12.5 {\rm km} \sim 3r_s$. In this paper, we set $R_*=3r_s$. In this paper, the inner boundary of our simulations is located at $3r_s$. Therefore, we can set the physical variables at the surface of a NS to be same as those at the inner boundary. We further assume that the black body emission photons at the surface of a NS are distributed spherically. The effective temperature ($T_*$) of the black body emission at the surface of a NS can be expressed as,
\begin{equation}
T_*=\left( \frac{L_*}{4\pi R_*^2 \sigma} \right)^{1/4}
\label{Tc}
\end{equation}
with $\sigma$ being the Stefan-Boltzmann constant. The soft photons generated at the surface of a NS can move outward radially and cool the accretion flow via inverse-Comptonization. The inverse-Compton cooling rate in Equation (4) can be expressed as follows (Sazonov et al. 2005),
\begin{equation}
Sc=4.1\times10^{-35}n^2(T_*-T_{\rm e})\xi
\label{comptoncooling}
\end{equation}
with $n$ and $T_{\rm e}$ being the number density of electrons and temperature of electrons, respectively. We define $\varrho=0.5$ and $m_p$ to be the mean molecular weight and the proton mass, respectively. The number density of electrons can be expressed as $n=\rho/\varrho m_p$. $\xi=L_* e^{-\tau_{es} (r)}/nr^2$ is the ionization parameter, with $\tau_{es}(r) =\int_0^r \rho \kappa_X dr$ being the X-ray scattering optical depth in radial direction. We set $\kappa_X=0.4 {\rm cm^2 g^{-1}}$.

For the accretion flow around a BH, we set $Sc=0$.

\subsection{Initial and boundary conditions}

The inner radial boundary of our simulations is located at $3r_s$. The outer radial boundary is located at $1000r_s$. As introduced above, for the simulations of HAFs around a BH, we set $r_s=2GM_{\rm BH}/c^2$. For the simulations of HAFs around a NS, we set $r_s=2GM_{\rm NS}/c^2$. Our simulation boundaries in $\theta$ direction are located at $\theta=0$ and $\theta=\pi$. We use $192 \times 88$ grids to resolve the computational domain. In order to have higher resolution close to the compact object, in the $r$ direction, the grids are logarithmically spaced. In $\theta$ direction, we put the grids uniformly.
At the inner and outer radial boundary, we use outflow boundary conditions. At the rotational axis, we use axis-symmetry boundary conditions.

The initial condition is a rotating torus with constant specific angular momentum embedded in a non-rotating low density medium.
%in our computational domain. The pressure and density in the torus are related through a polytropic equation of state $p=A\rho^\gamma$. The structure of the torus is given by (Papaloizou \& Pringle 1984):
%\begin{equation}
%\frac{p}{\rho}=\frac{GM}{(n+1)R_0}\left[ \frac{R_0}{r}-\frac{1}{2}\left(\frac{R_0}{r\sin\theta}\right)-\frac{1}{2d} \right]
%\end{equation}
%In this equation, $M$ is the mass of the central object. In simulations of NS HAF, $M=M_{\rm NS}$. In simulations of BH HAF, $M=M_{\rm BH}$. $R_0$ is the location of torus center.  $R_0$ is located at 470 $r_s$. For the black hole HAF simulations, $R_0=940GM_{\rm BH}/c^2$; for the NS HAF simulations $R_0=940GM_{\rm NS}/c^2$.  $n=(\gamma-1)^{-1}$ is the polytropic index. $d=1.125$ is the distortion parameter of the torus.
The exact formula for the torus can be found in Papaloizou \& Pringle (1984, see also Stone et al. 1999). The torus center (or density maximum) is located at $300r_s$. The maximum density at the torus center is $\rho_{\rm max}$ (see Table 1 for the values of $\rho_{\rm max}$ in different models).

\begin{figure*}
\begin{center}
\includegraphics[scale=0.7]{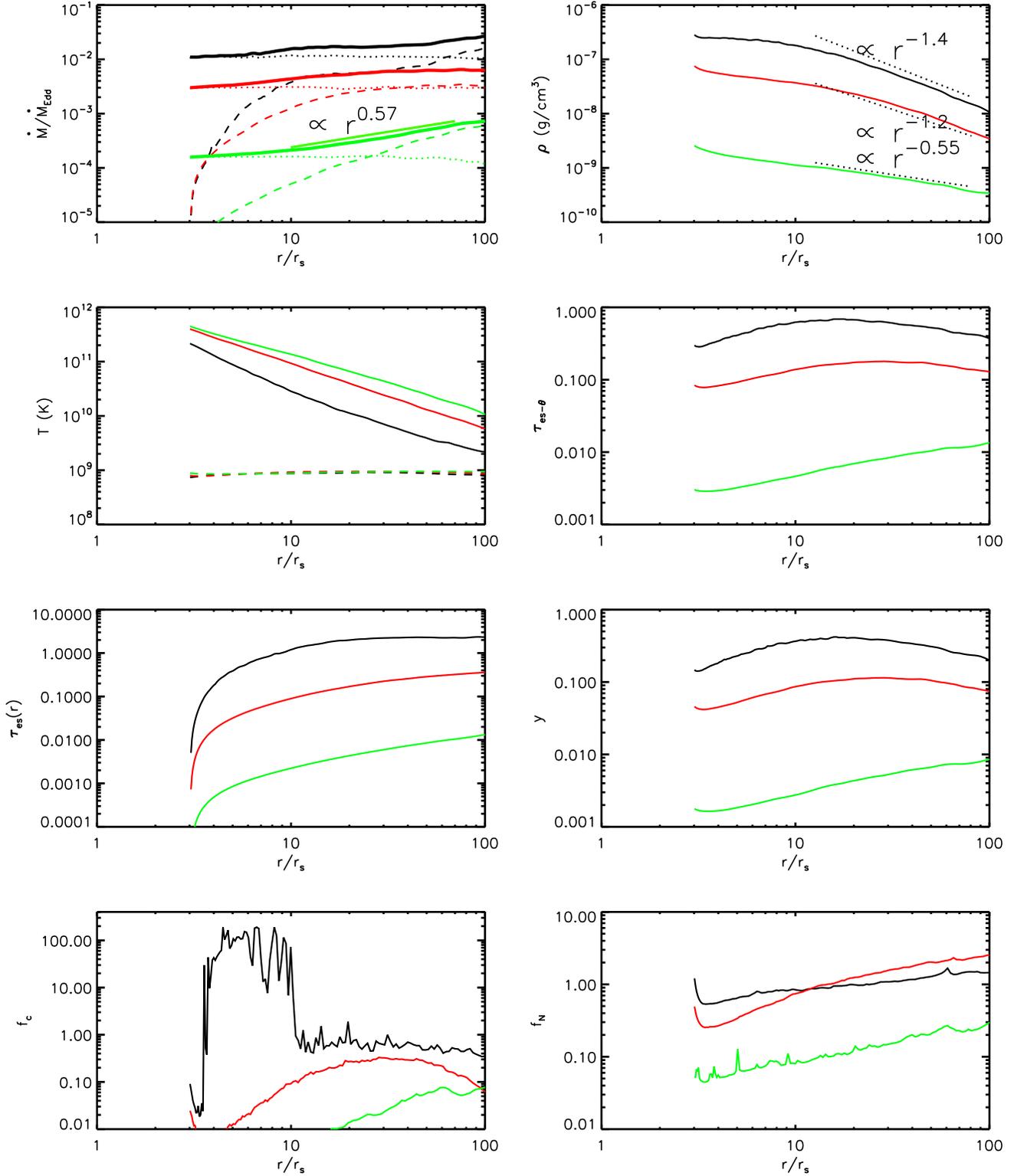}\hspace*{0.7cm}
\hspace*{0.5cm} \caption{Time-averaged (from $t=2$ to 4 orbital time measured at $100r_s$) variables for models NS1 (black lines), NS2 (red lines) and NS3 (green lines). In the top left panel, the radial profiles of mass inflow (solid lines, Equation (\ref{inflowrate})), outflow (dashed lines, Equation (\ref{outflowrate})) and net rates are plotted. The short green line is a radial power law function fit to the mass inflow rate profile outside $10r_s$ for model NS3. The top right panel shows the radial profiles of the volume-weighted density. The three dotted lines are radial power law function fits to the results. The left panel in the second row shows the density-weighted temperatures for ions (solid lines) and electrons (dashed lines). The right panel in the second row shows the electron scattering optical depth in $\theta$ direction. The left panel in the third row shows the density-weighted electron scattering optical depth in $r$ direction. The right panel in the third row shows the radial profiles of the Compton $y-$parameter. The bottom left panel shows the density-weighted radial profiles of the ratio of $Q^-$ to the viscous heating rate (Equation (\ref{coolratio})). The bottom right panel shows the density-weighted radial profiles of the ratio of $S_c$ to the viscous heating rate (Equation (\ref{coolratio2})).  \label{Fig:NS}}
\end{center}
\end{figure*}

\begin{figure}
\begin{center}
\includegraphics[scale=0.5]{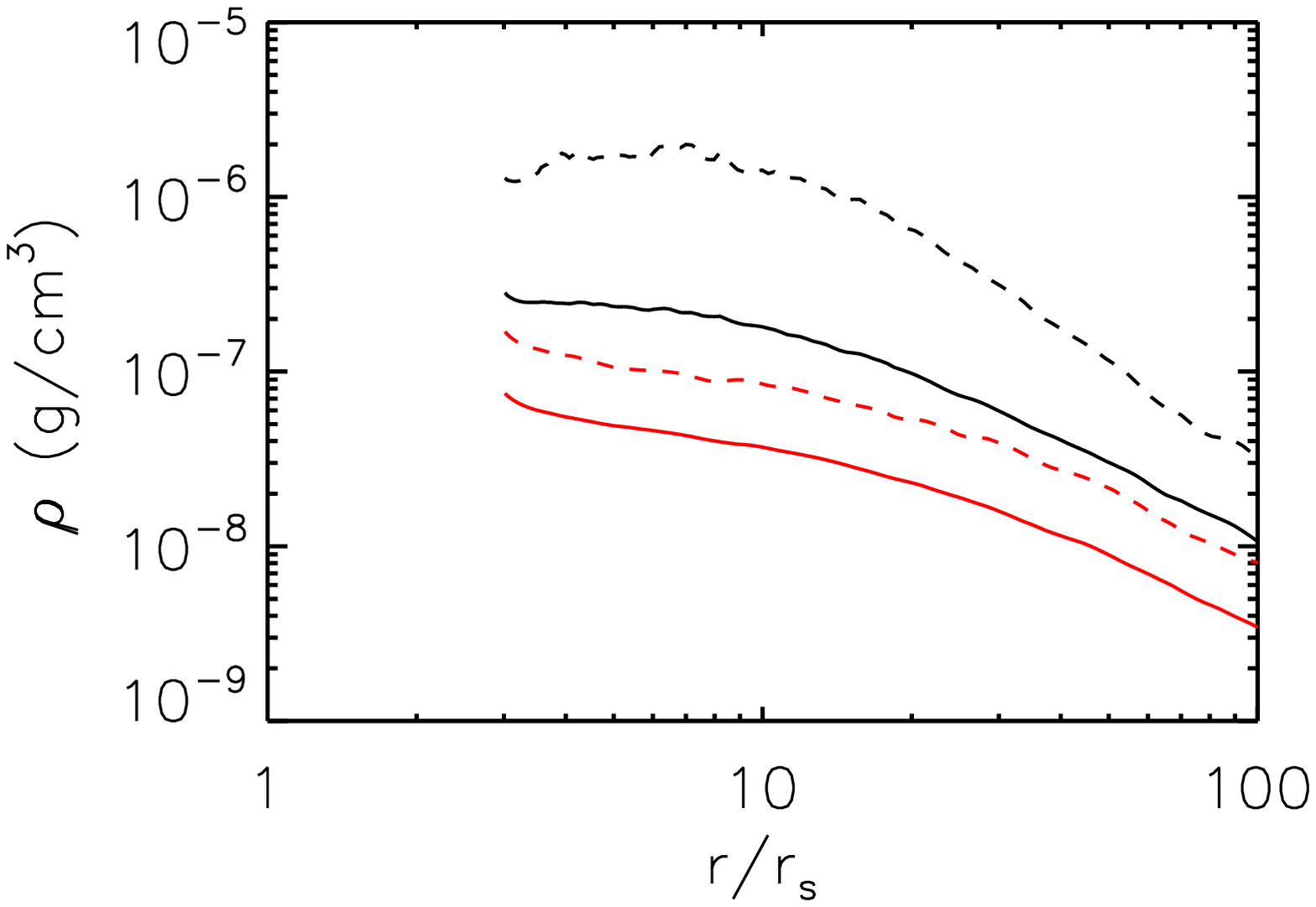}\hspace*{0.3cm}
\hspace*{0.5cm} \caption{Time-averaged (from $t=2$ to 4 orbital time measured at $100r_s$) density for models NS1 (black lines), NS2 (red lines). The dashed lines show the density at the equatorial plane. The solid lines show the $\theta$ angle averaged (from $\theta=0-\pi$) density. The $\theta$ angle averaged densities are same as those shown in the top-right panel of Figure \ref{Fig:NS}.  \label{Fig:dclumps}}
\end{center}
\end{figure}

\section{Results}
\begin{figure*}
\begin{center}
\includegraphics[scale=0.7]{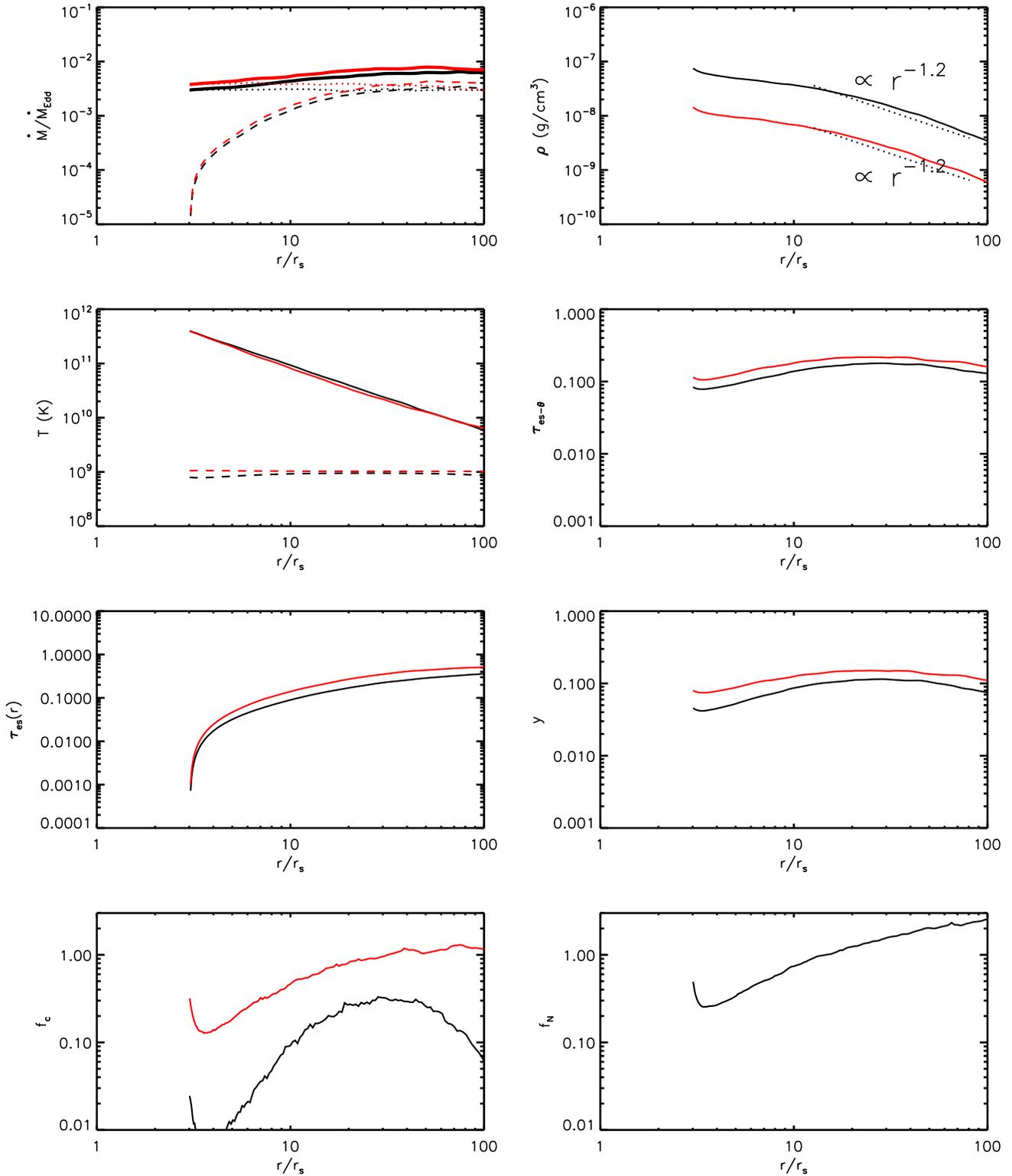}\hspace*{0.7cm}
\hspace*{0.5cm} \caption{Same as Figure \ref{Fig:NS} but for models NS2 (black lines) and BH2 (red lines). \label{Fig:NSvsBH1}}
\end{center}
\end{figure*}

In the present paper, time is expressed in unit of orbital time at 100$r_s$. We first run the simulations without radiative cooling to $t=1.5$ orbits. At $t=1.5$ orbits, the flow has achieved a quasi-steady state. We turn on radiative cooling at $t=1.5$ orbits. We find that a new quasi-steady state can be achieved at $t=2$ orbits. The time-averaged results presented in this paper are obtained by averaging the simulation data between $t=2$ to $4$ orbits. In this paper, we use ``quasi-steady state" instead of ``steady state". Actually, the accretion flow is always very turbulent if we see the snapshot of the physical variables as shown by previous works (e.g. Stone et al. 1999). The accretion flow can achieve a quasi-steady state in the time-averaged sense. One usually adopted diagnostic for judging whether the flow has achieved a quasi-steady state is the time-averaged net mass accretion rate (see the dotted lines in the top-left panels of Figures \ref{Fig:NS}, \ref{Fig:NSvsBH1} and \ref{Fig:NSvsBH2}). If the time-averaged net mass accretion rate is almost a constant of radius, one usually believes that the accretion flow has achieved a quasi-steady state. From the plotted net accretion rate in Figures \ref{Fig:NS}, \ref{Fig:NSvsBH1} and \ref{Fig:NSvsBH2}, we can see that all the models in this paper have achieved a quasi-steady state.
We take model NS1 to illustrate the depletion timescale of the torus. In this model, the mass of the torus is $4.2 \times 10^{16} {\rm g}$. The mass accretion rate in this model is $2.2 \times 10^{16} {\rm g/s}$. Therefore, the depletion timescale of the tours is $\sim 2$ seconds. The orbital timescale at $100r_s$ is $1.98 \times 10^{-2}$ seconds. Therefore, in order to deplete the torus, we need 100 orbital times at $100r_s$. In this paper, we obtain the time-averaged flow properties from 2-4 orbital times at $100r_s$. Therefore, the properties of the flow obtained in this paper are safe and not be affected by the depletion of the torus. We summarize the models in Table 1.

The equations calculating mass inflow and outflow rates are same as those in Stone et al. (1999). They are as follows,
(1) inflow rate (or accretion rate)
\begin{equation}
\dot {M}_{\rm in} (r)=-2\pi r^2 \int_{\rm 0}^{\rm \pi}
\rho \min (v_r, 0) \sin\theta d\theta
\label{inflowrate}
\end{equation}
(2) outflow rate
\begin{equation}
\dot {M}_{\rm out} (r)=2\pi r^2 \int_{\rm 0}^{\rm \pi}
\rho \max (v_r, 0) \sin\theta d\theta
\label{outflowrate}
\end{equation}
Adding Equations (\ref{inflowrate}) and (\ref{outflowrate}), we can obtain the net accretion rate.

For some two-dimensional variables (e.g., temperature, the ratio of radiative cooling rate to the viscous heating rate (Equation \ref{coolratio}) below), we can obtain their one-dimensional ($r$ direction) profiles by density-weighted them in the $\theta$ direction as follows,
\begin{equation}
Va (r)=\frac{2\pi r^2 \int_0^\pi Va(r, \theta) \rho (r, \theta) \sin \theta d \theta}{2\pi r^2 \int_0^\pi \rho (r, \theta) \sin \theta d \theta}
\label{density-weight}
\end{equation}
In Equation (\ref{density-weight}), $Va (r, \theta)$ denotes the two-dimensional physical variable, $Va(r)$ denotes the density-weighted one-dimensional variable.

\subsection{Accretion flows around neutron star}
In the accretion flow around a NS, one portion of the viscous heating energy ($Q_{\rm vis}$) is stored in gas as entropy and advected to the NS. The second portion of the viscous heating energy is lost by radiation of electrons ($Q^-$). The last portion of the viscous heating energy is lost through the inver-Compton scattering cooling (the term $S_c$ in Equation (4)) via collision between the soft photons from the surface of the NS and the accretion flow. We define the ratio of $Q^-$ to the viscous heating rate as,
\begin{equation}
f_c=\frac{Q^-}{Q_{\rm vis}}
\label{coolratio}
\end{equation}
We define the ratio of $S_c$ to the viscous heating rate as,
\begin{equation}
f_N=\frac{S_c}{Q_{\rm vis}}
\label{coolratio2}
\end{equation}

The radial profiles of $f_c$ and $f_N$ are shown in the bottom left and bottom right panels of Figure \ref{Fig:NS}, respectively. Compared to models NS1 and NS2, the mass accretion rate in model NS3 is lowest (see the top-left panel of Figure \ref{Fig:NS}). In this model, the accretion rate is low, radiative cooling is not important. The cooling term ($Q^-$) is more than one order of magnitude smaller than the viscous heating term. Although most of the viscous heating energy is advected to the surface of the NS and emitted out there, the emitted thermal soft photons have very low interaction efficiency with the accretion flow ($f_N \ll 1$). The reason is as follows. The inverse-Compton cooling (Equation (\ref{comptoncooling})) $S_c \propto n^2 L_*/n \propto n^2 \dot M/n $. $\dot M$ is proportional to $n$. Therefore, we have $S_c \propto n^2$. In model NS3, the very low density results in a very low interaction efficiency between soft photons from the surface of the NS and the accretion flow. The dynamical properties of the HAF in model NS3 are quite similar as those of a non-radiative HAF around a BH. For example, the mass inflow rate decreases inwards (see the top left panel of Figure \ref{Fig:NS}). The mass inflow rate in the region $10r_s<r<100r_s$ can be described as a power law function of radius $\dot M_{\rm in} \propto r^s$, with $s=0.57$. Numerical simulations of non-radiative HAFs around BHs also find that mass inflow rate outside $10r_s$ can be described as a power-law function of radius, with the power-law index in the range $0.5<s<1$ (e.g., Stone et al. 1999; Sadowski et al. 2013; Yuan et al. 2012, 2015; Bu et al. 2013). As explained in previous works (e.g., Sadowski et al. 2013; Yuan et al. 2012, 2015; Bu et al. 2013; Gu 2015; Almeida \& Nemmen 2019), the inward decrease of mass inflow rate is due to the presence of outflow/wind (green dashed line in top-left panel of Figure \ref{Fig:NS}). The presence of wind generated by a HAF has been confirmed by observations (e.g., Crenshaw \& Kraemer 2012; Wang et al. 2013; Cheung et al. 2016; Homan et al. 2016; Almeida et al. 2018; Ma et al. 2019; Park et al. 2019). If the accretion rate is smaller than that in model NS3, the values of $f_c$ and $f_N$ will be smaller, and the dynamical properties of the HAF around a NS will be same as those of a non-radiative HAF around a BH.

In model NS2, the accretion rate (or density) is higher than that in model NS3. For hot accretion flow, viscous heating rate is proportional to $n$ (see the second term on the right hand side of Equation (3)). With the increase of accretion rate (or density), the radiative cooling rate increases faster than viscous heating rate. For example, the bremsstrahlung cooling rate $q_{\rm brem}^-
\propto n^2$. With the increase of accretion rate, the inverse Compton cooling ($S_c$) rate also increases faster than viscous heating rate because $S_c \propto n^2$. From the bottom-left panel, we can see the value of $f_c$ in model NS2 is much bigger than that in model NS3. However, in model NS2, $f_c$ is still smaller than 1. The value of $f_N$ is smaller than 1 inside $15r_s$. Outside $15r_s$, $f_N>1$. In model NS2, the invser-Compton cooling rate due to the presence of the thermal soft photons from the surface of the NS dominates viscous heating rate outside $15r_s$. In model NS1, the value of $f_c$ in the region $3.2r_s < r < 10r_s$ is significantly larger than 1. In the region $10r_s < r < 20r_s$, the value of $f_c$ can also be larger than 1. Outside $20r_s$, the value of $f_N$ is much larger than 1. Therefore, in model NS1, almost at the whole radii, the cooling rate is larger than the viscous heating rate. The parameter $f_N=S_c/Q_{\rm vis} \propto n^2 \exp ({-\tau_{es} (r))}/n \propto n \exp{(-\tau_{es}(r))}$. The density-weighted radial profiles of $\tau_{es} (r)$ are shown in the left panel of the third row. In the region $r < 10r_s$, the values of $\tau_{es} (r)$ in all models are much smaller than 1. Therefore, the value of $f_N$ in this region mainly depends on $n$. In the region $r < 10 r_s$, the value of $f_N$ in model NS1 is much bigger than that in model NS2. Outside the region $10r_s$, the electron scattering optical depth in radial direction ($\tau_{es} (r)$) in model NS1 is significantly larger than that in model NS2. The larger value of $\tau_{es}(r)$ in the region outside $10r_s$ in model NS1 (compared to that in model NS2) makes the value of $f_N$ in this region much smaller than that in the same region in model NS2.

\begin{figure*}
\begin{center}
\includegraphics[scale=0.7]{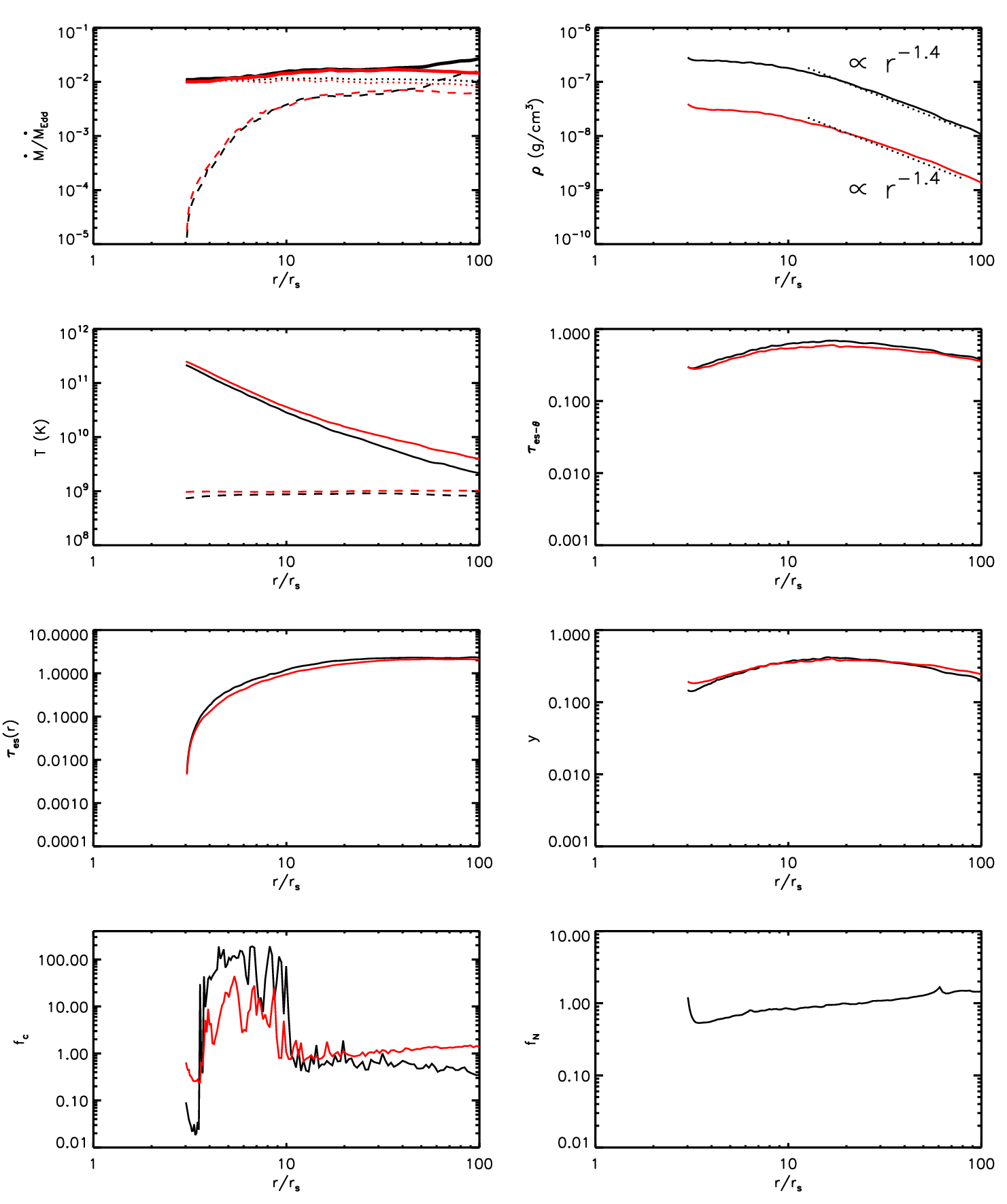}\hspace*{0.7cm}
\hspace*{0.5cm} \caption{Same as Figure \ref{Fig:NS} but for models NS1 (black lines) and BH1 (red lines). \label{Fig:NSvsBH2}}
\end{center}
\end{figure*}

In model NS1, there is a sharp jump of $f_c$ at $\sim 10 r_s$. The sharp jump is not a numerical noise. Yuan (2001) analytically studied the properties of hot accretion flow. In his highest accretion rate model, it is found that outside $10 r_s$, the value of $f_c$ increases inwards. It is also found that outside $10r_s$, $f_c$ is $\sim 1$. A sharp jump of $f_c$ at $ r \sim 10 r_s$ is also found (see the right panel of Fig.2 in Yuan 2001). The sharp jump behavior of $f_c$ in models NS1 and BH1 (see the bottom-left panel of Figure \ref{Fig:NSvsBH2}) is quite similar as that found by Yuan (2001). The physical reason for the sharp jump of $f_c$ at $ r \sim 10r_s$ is as follows. In the region $r<10r_s$, cooling is significantly stronger than heating, the gas is thermally unstable and high density cold clumps form around the equatorial plane. In Figure \ref{Fig:dclumps}, we plot the density at the equatorial plane and the $\theta$ angle averaged (or volume weighted) density for models NS1 and NS2. For model NS1, in the region $r<10 r_s$, the density at equatorial plane is 10 times higher than the $\theta$ angle averaged density. The presence of high density cold clumps at equatorial plane makes the high contrast between density at equatorial plane and the $\theta$ angle averaged density. For a comparison, in model NS2, no cold clumps form at equatorial plane. Therefore, in model NS2, the density at equatorial plane is just 2 times higher than the $\theta$ angle averaged density. In model NS1, the presence of high density cold clumps at the equatorial plane in the region $r<10 r_s$ makes the radiation in this region significantly stronger. This is the reason why there is a sharp jump of $f_c$ at $r \sim 10 r_s$. We note that in Pringel et al. (1973), it is anticipated that when the accretion rate reaches a high value, cold clumps will form in the inner region close to the central object.

For an accretion flow, the internal energy equation is $de/dt+pdv=dQ=Q_{\rm vis}-Q_{\rm cool}$. In this equation, $e$ is gas internal energy per volume, $-pdv$ is compression work heating rate, $Q_{\rm cool}$ is the radiative cooling rate. In the accretion flow around a NS, $Q_{\rm cool}=Q^-+Sc$. In the accretion flow around a BH, $Q_{\rm cool}=Q^-$. We can rewrite this equation as $de/dt=dQ-pdv=Q_{\rm vis}-pdv-Q_{\rm cool}$. In this equation, the term $dQ$ ($=Q_{\rm vis}-Q_{\rm cool}$) can be positive, zero or negative. It is not violating energy conservation law that if we have $dQ$ to be negative. If the sum of the viscous heating and compression work heating terms ($Q_{\rm vis}-pdv$) is smaller than radiative cooling term, it is also not violating the energy conservation law. In this case it means that with time evolution, the internal energy of gas will decrease. For the classic hot accretion flow model (e.g., Narayan \& Yi 1994), accretion rate is extremely low, the term $dQ$ is positive. With the increase of accretion rate, $dQ$ can become negative. If $dQ$ is negative and $dQ-pdv$ is positive, radiative cooling rate is smaller than the sum of viscous heating rate and compression work heating rate, the accretion flow is ``Luminous hot accretion flow" (LHAF, Yuan 2001, see also Yuan \& Narayan 2014). In this case, the accretion flow can still be hot. The word ``luminous" means that the luminosity of LHAF is much higher than that of the classic hot accretion flow. In black hole X-ray binaries, the hard state often reaches luminosities $\sim 10\%$  of Eddington luminosity during hard-to-soft transition. Yuan \& Zdziarski (2004) suggested that such luminous hard state systems, and also some Seyfert galaxies, may be explained by the LHAF model. This has been confirmed in detailed modeling of XTE J1550-564 (Yuan et al. 2007), where the X-ray spectrum is naturally explained by the LHAF model. If the accretion rate increases to a even higher value, the term $dQ-pdv$ can be negative, in this case, hot accretion flow solution will not exist. For model NS2, we average for the whole accretion flow and find that the ratio of cooling rate to heating rate is 0.6. For model NS1, this ratio is 0.84.

The temperature of ions is determined by the viscous heating rate and the Coulomb collision cooling rate (see Equation (3)). Because the Coulomb collision cooling rate $q^{\rm ie} \propto n^2 \propto \dot M^2$, with the increase of accretion rate, Coulomb collision cooling rate increases faster than viscous heating rate. Therefore, the temperature of ions decreases with the increase of mass accretion rate (see the left-panel of the second row in Figure \ref{Fig:NS}). The temperature of electrons almost does not depend on mass accretion rate. In the radial self-similar solution of accretion flows around a NS, Qiao \& Liu (2018) also find that temperature of electrons is not sensitive to mass accretion rate. The reason is that for electrons at any mass accretion rate, the coulomb collision heating energy can be very efficiently lost by the radiative cooling. The temperature of electrons obtained in this paper is $2-3$ times higher than what is observed in a NS accretion flow (Burke et al. 2017). The reason may be as follows. The temperature of electrons depends on heating and cooling of electrons. There are two heating sources for electrons. The first one is the energy transfer from ions to electrons by Coulomb collisions, which has been appropriately included in our simulations. The second one is the directly heating by viscous stress. In other words, one portion of viscous heating energy can go directly into electrons. Because, the exact fraction of viscous heating energy that directly goes into electrons is still unclear and under debate (see Yuan \& Narayan 2014 for a review), in this paper, following Narayan \& Yi (1995), we assume that all the viscous hearting energy goes into ions. In a realistic NS accretion system, may be a portion of viscous heating energy directly goes into electrons. In this case, because electrons can very quickly radiate away energy they receive, we may expect a higher radiative efficiency and more energy is lost from the accretion system by radiation. Correspondingly, the temperature of electrons in the realistic NS accretion system should be lower.

The gas pressure gradient force (per unit mass) which drives outflow/wind is $-\nabla p/\rho \propto T$. In models NS2 and NS3, the temperature of ions is lower (compared to that in model NS3), the outflow/wind (compared to inflow) is weaker than that in model NS3. In model NS3, the outflow rate exceeds the net accretion rate significantly outside $25r_s$. In models NS2 and NS3, the outflow rate just roughly equals to the net rate outside $25r_s$. Therefore, in models NS2 and NS3, the mass inflow rate outside $20r_s$ is almost a constant with radius.

In the top-right panel of Figure \ref{Fig:NS}, we plot the volume weighted radial profiles of density.
\begin{equation}
\rho (r) = \frac{2 \pi r^2 \int_{\rm 0}^{\pi} \rho \sin \theta d\theta}{2 \pi r^2 \int_{\rm 0}^{\pi} \sin \theta d\theta}
\label{density}
\end{equation}
The three dotted lines are radial power law function fits to the density profiles. In model NS3, strong outflow/wind can take away mass, therefore, the density increases slower inwards. The radial profile of density can be described as $\rho \propto r^{-0.55}$. The result is consistent with that found in non-radiative simulations of HAF around a BH (e.g., Stone et al. 1999). In models NS1 and NS2, the outflow is weak (compared to inflow). The density increases faster inwards. The radial profile of density is $\rho \propto r^{-p}$. The values of $p$ are 1.4 and 1.2 for models NS1 and NS2, respectively. For a HAF without wind, we will have $\rho \propto r^{-1.5}$ (e.g., Narayan \& Yi 1994). The radial density profiles in models NS1 and NS2 are quite similar as that in models without wind.

The electron scaterring optical depth in vertical direction $\tau_{es-\theta}$ is an important parameter for calculating the spectrum of a HAF (Burke et al. 2017). It is defined as follows,
\begin{equation}
\tau_{es-\theta}=\int_{\rm 0}^{\rm \pi} \rho \kappa r d\theta
\label{tao}
\end{equation}
In the right panel of the second row of Figure \ref{Fig:NS}, we plot the radial profiles of $\tau_{es-\theta}$. As expected, the value of $\tau_{es-\theta}$ increases with the increase of mass accretion rate.

The other parameter that affecting the spectrum of a HAF is the Compton $y-$parameter (Burke et al. 2017). It is defined as follows,
\begin{equation}
y=\frac{4 k T_e}{m_e c^2} \max(\tau_{es-\theta}, \tau_{es-\theta}^2)
\end{equation}
where $k$ and $m_e$ are Boltzmann constant and mass of electron, respectively. The temperature of electrons is almost independent of mass accretion rate (see the left-panel in the second row of Figure \ref{Fig:NS}). Therefore, the value of $y$ increases with mass accretion rate (see the right-panel in the second row of Figure \ref{Fig:NS}), because $\tau_{es-\theta}$ increases with mass accretion rate.

\subsection{Comparison with one-temperature simulation}
In Paper I, we assume that electrons and ions have same temperatures and perform simulations of HAF around NS. Also, in Paper I, the radiative cooling term $Q^-$ (bremsstrahlung, synchrotron and their self-Comptonization) is neglected. In paper I, we find that the soft photons from the surface of a NS can cool the gas via Comptonization. We find that for the one-temperature HAF, outflows are completely absent when $\dot m>10^{-4}$. However, in the present paper, we find that only when $\dot m>10^{-3}$, can outflows be significantly weak (models NS1 and NS2). In model NS3, with $\dot m=1.6\times 10^{-4}$, outflows are very strong. We define a critical accretion rate ($\dot m_c$) as the accretion rate above which outflows are significantly weak compared to inflows. The value of $\dot m_c$ in the present two-temperature simulations is one order of magnitude higher than that in the one-temperature simulations in Paper I. The reason is as follows. In the one-temperature simulations (Paper I), the soft photons from the surface of a NS can very effectively cool the electrons via Comptonization when $\dot m>10^{-4}$. Because in Paper I, we assume that ions have same temperature as electrons, therefore, ions are also have very low temperature. In the two-temperature simulations, the soft photons from the surface of a NS can only cool the electrons. The temperature of ions can still be very high even temperature of electrons is low. For example, in Paper I, when $\dot m=1.2\times 10^{-4}$, the gas (ions and electrons) temperature at the inner radial boundary is $5\times 10^{10}$ K. In model NS3 with $\dot m=1.6\times 10^{-4}$ in the present paper, the ions temperature at the inner radial boundary is $5\times 10^{11}$ K. The gas pressure gradient force in model NS3 in the present paper can be one order of magnitude higher than that in the model with $\dot m=1.2\times 10^{-4}$ in Paper I. This is the reason why the critical accretion rate $\dot m_c$ in the present paper is significantly higher than that in the one-temperature simulations in Paper I.

\subsection{Black hole versus Neutron star}
In models BH1, BH2, and BH3, the central object is a BH. All of the energy stored in the accretion flow will eventually go into the BH horizon. The cooling term $Sc=0$ in models BH1, BH2, and BH3.

In the black hole accretion flow, one portion of the viscous heating energy is stored in gas as entropy and advected into the BH horizon. The other portion is lost by radiation of electrons (the term $Q^-$). In model BH3 with the lowest accretion rate, we find that more than $80\%$ of the viscous heating energy is stored in gas and advected into the black hole horizon; radiative cooling is negligible. Therefore, in model BH3, the dynamical properties of the HAF are quite similar as those of a non-radiative HAF. As mentioned above, for the accretion flow around a NS, when the accretion flow has $\dot m < \sim 10^{-4}$ (model NS3), the dynamical properties of the HAF are also quite similar as those of a non-radiative HAF due to the very low cooling rate. Therefore, we can conclude that when $\dot m $ is smaller than $\sim 10^{-4}$, the dynamical properties of a HAF around a NS are quite similar as those of a HAF around a BH. For example, we find that in model BH3, the radial profile of the mass inflow rate outside 10$r_s$ can be described as $\dot M_{\rm in} \propto r^s$, with $s=0.57$. In model NS3, the value of $s$ is also 0.57. In model BH3, the density profile can be described as $\rho \propto r^{-p}$, with $p=0.55$. In model NS3, the value of $p$ is also 0.55.

We now compare the properties of the HAF in model NS2 to those of the HAF in model BH2. The values of $\dot m$ at the inner boundary in these two models are very similar. The results are shown in Figure \ref{Fig:NSvsBH1}. From the bottom two panels of Figure \ref{Fig:NSvsBH1}, we can see that in model NS2, the dominant cooling process is the Comptonization cooling ($Sc$) by collisions of electrons and the soft photons from the surface of the NS. In model BH2, the term $Sc=0$. However, the value of $f_c$ is much larger in model BH2 than that in model NS2. The gas density in NS2 is much higher than that in model BH2, one may expect that the value of $f_c$ in model NS2 should be higher than that in model BH2. However, the trend is opposite. The reason is as follows. The exact formula of viscous heating rate is expressed in Equation (3). The viscous heating rate is proportional to $\rho/M (r/r_s)^{-2.5}$, where $M$ is mass of the central object. For simplicity, we assume that the cooling is bremsstrahlung radiation. The cooling rate is proportional to $\rho^2 T_e^{0.5}$. We have, $f_c \propto \rho T_e^{0.5} (r/r_s)^{2.5} M$. At a fixed radius (in unit of Schwarzschild radius), the temperatures of electrons in models NS2 and BH2 are almost same. The density in model BH2 is roughly 5 times smaller than that in model NS2. However, the black hole mass in BH2 is roughly 7 times higher than the Neutron star mass in model NS2. Therefore, we would expect a slightly higher value of $f_c$ in model BH2 than that in model NS2. In reality, if we consider all the radiative cooing process (bremsstrahlung, synchrotron and their Comptonization), the situation will be more complex. It is hard to expect that with a similar accretion rate (in unit of Eddington rate), in which model (BH or NS accretion flow), the value of $f_c$ will be larger. In model BH2, inside $\sim 20 r_s$, $f_c \sim 0.5$ and outside $20r_s$, the value of $f_c \sim 1$. Inside $\sim 20r_s$, about $50\%$ of the viscous heating energy is lost by radiative cooling ($Q^-$). Outside $20r_s$, all of the viscous heating energy is lost by radiative cooling ($Q^-$). In model NS2, the term $Sc$ is the dominant cooling term. In model NS2, inside $\sim 20 r_s$, $f_N \sim 0.5$ and outside $20r_s$, the value of $f_N > 1$. In model NS2, inside $\sim 20r_s$, about $50\%$ of the viscous heating energy is lost by the cooling term $Sc$. Outside $20r_s$, all of the viscous heating energy is lost by the cooling term $Sc$. From the comparisons we can see that the fraction of viscous heating energy that lost by cooling in model NS2 is very similar as that in model BH2. In other words, the fraction of viscous heating energy that stored in gas as entropy in model NS2 is very similar as that in model BH2. The properties of the HAF in model NS2 are quite similar as those in model BH2. The absolute vale of density in model NS2 is higher than that in model BH2 (see the top right panel in Figure \ref{Fig:NSvsBH1}). The reason is as follows. The mass of the BH is much larger than that of the NS. The values of $\dot m$ in the two models are quite similar. Therefore, the absolute value of density in model NS2 is much higher than that in model BH2. However, the slope of the density profile in model NS2 is almost same as that of the density profile in model BH2. In both models of BH2 and NS2, outflow is weak and the radial profile of mass inflow rate is quite flat. The temperature of ions in model NS2 is almost same as that in model BH2. The temperature of electrons in model NS2 is just lightly lower than that in model BH2. The electron scattering optical depths ($\tau_{es}(r)$ and $\tau_{es-\theta}$) in both $r$ and $\theta$ directions in model BH2 are also slightly larger than those in model NS2. The Compton $y$ parameter in model BH2 is slightly higher.

Figure \ref{Fig:NSvsBH2} shows the comparisons of models NS1 and BH1. The values of $\dot m$ at the inner boundary in these two models are very similar. From the bottom panels we can see that inside $10r_s$ in both models, the value of $f_c$ is significantly larger than 1. Therefore, in this region, radiative cooling ($Q^-$) dominates the viscous heating. Also, in this region ($r<10r_s$), the ratios of the cooling rate to viscous heating rate in the two models are very similar. Outside $10r_s$, the ratios of cooling rate to the heating rate in the two models are also very similar. For example, in model BH1, at $100r_s$, the ratio of cooling rate to viscous heating rate $f_c \sim 1.5$; in model NS1, at the same radius, this ratio is $f_c+f_N \sim 1.8$. The properties of the accretion flow in model NS1 are also quite similar as those in model BH1. The Comptonized spectrum is mainly determined by the Compton $y$ parameter, and is slightly affected by the temperature of the seed photons. The Compton $y$ parameter in model NS1 is almost same as that in model BH1. Therefore, we can expect that the shape of the Comptonized spectrum (or hard X-ray spectrum) in model NS1 should be almost same as that in model BH1. However, in the NS accretion flow spectrum, there is a peak in the soft X-ray band due to the presence of soft photons from the surface of the NS. In the BH accretion flow spectrum, there is no such peak in the soft X-ray band. In this sense, the spectrum of a NS system is different from that of a BH system.

From Figures \ref{Fig:NSvsBH1} and \ref{Fig:NSvsBH2}, we can see that the BH models and NS models have almost same electron temperature. Usually, it is expected that electron temperature of a NS accretion flow is much lower than that of a BH accretion flow. The reason is that, the soft photons from the surface of a NS can effectively cool the accretion flow. For a BH accretion flow, there is no such soft photons. Whether is this expectation valid depends on accretion rate. For low accretion rate systems, this expectation is valid. The reason is as follows. For low accretion rate BH system (e.g. model BH3), the radiation of electrons (bremsstrahlung, synchrotron and their Comptonization) is not so important, the electrons have higher temperature. For a NS system with a similar accretion rate (e.g. model NS3), the radiation of electrons (bremsstrahlung, synchrotron and their Comptonization) is also not so important. However, the soft photons from the surface of the NS can effectively cool electrons and make temperature of electrons lower. We find that the electron temperature in model NS3 is two times lower than that in model BH3.  For high accretion rate system, the expectation that NS system has a much lower electron temperature than BH system is not valid. The reason is as follows. In the high accretion rate BH system (e.g., models BH1 and BH2), the radiation of electrons (bremsstrahlung, synchrotron and their Comptonization) is very strong and the radiative efficiency is very high.  In a NS system with a similar accretion rate, the radiation of electrons (bremsstrahlung, synchrotron and their Comptonization) is also very strong and the presence of soft photons from the surface of a NS cannot significantly affect the radiative efficiency of the accretion flow. This is the reason why for the high accretion rate NS and BH systems (models NS1, NS2, BH1 and BH2), the temperatures of electrons are quite similar.

\section{Summary}
Numerical simulations of two-temperature radiative hot accretion flows around NSs are performed. We find that when $\dot m < \sim 10^{-4}$, the cooling of the HAF (terms $Q^-$ and $Sc$) is not important and outflows are very strong. The dynamical properties of the HAF are same as those of a non-radiative HAF. When $\dot m$ is increased to be above $\sim 10^{-3}$, cooling becomes important and outflows become significantly weak. In the range $10^{-4} < \dot m < 10^{-3}$, the HAFs transients from a strong outflow phase to a very weak outflow phase with increase of $\dot m$. When $\dot m \sim 10^{-3}$, the cooling due to the Compton scattering between thermal soft photons from the surface of the NS and the electrons is much larger than the term $Q^-$. When $\dot m$ is increased to $1.1\times 10^{-2}$, inside $10r_s$, the term $Q^-$ is much larger than the term $Sc$. Outside $10r_s$, the term $Sc$ dominates.

We also compare the properties of the HAF around a NS to those of the HAF around a BH. We find that with a similar accretion rate (in unit of Eddington rate), the dynamical properties of the HAF around a NS are quite similar as those of the HAF around a BH. However, we note that the emitted spectrum of a HAF around a NS can be quite different from that of a HAF around a BH. For the HAF around a NS, in addition to the photons directly radiated out by the accretion flow (the term $Q^-$), we can also observe a soft X-ray photon flux coming from the surface of the NS. Therefore, in the spectrum of a NS-LMXB system, there is a thermal soft X-ray component. This point has been confirmed by observations (e.g., Jonker et al. 2004; Armas Padilla et al. 2013a, 2013b; Degenaar et al. 2013b; Campana et al. 2014). For the BH-LMXB system, such a thermal soft X-ray component is not observed.

\section*{Acknowledgments}
We thank Yuan F. for useful discussions. Bu D. thanks Gan Z. for modifying the two-temperature cooling part of the simulation code. Bu D. is supported in part by the Natural Science Foundation of China (grants  11773053, 11573051, 11633006 and 11661161012) and the Key
Research Program of Frontier Sciences of CAS (No. QYZDJSSW-
SYS008). Qiao E. is supported by the Natural Science Foundation of China (grant 11773037), the Strategic Pioneer Program on Space Science, Chinese Academy of Sciences (grant XDA15052100), the gravitational wave pilot B (grant XDB23040100), and the National Program on Key Research and Development Project (grant 2016YFA0400804). Yang X. is supported by the Natural Science Foundation of China (grant 11973018) and Natural Science Foundation Project of CQ CSTC (grant cstc2019jcyj-msxmX0581). This work made use of the High Performance Computing Resource in the Core Facility for Advanced Research Computing at Shanghai Astronomical
Observatory.


\begin{thebibliography}{99}
\bibitem[\protect\citeauthoryear{}{}]{} Almeida, I, Nemmen, R., Wong, K. et al. 2018, MNRAS, 475, 5398
\bibitem[\protect\citeauthoryear{}{}]{} Almeida, I, \& Nemmen, R. 2019 (arXiv:1905.13708)
%\bibitem[\protect\citeauthoryear{}{}]{} Avara, M. J., McKinney, J. C., \& Reynolds, C. S. 2016, MNRAS, 462, 636
\bibitem[\protect\citeauthoryear{}{}]{} Armas Padilla, M., Degenaar, N., \& Wijnands, R. 2013a, MNRAS, 434, 1586
\bibitem[\protect\citeauthoryear{}{}]{} Armas Padilla, M., Wijnands, R., \& Degenaar, N. 2013b, MNRAS, 436, L89
\bibitem[\protect\citeauthoryear{}{}]{} Bahramian, A., Heinke, C. O., Sivakoff, G. R., et al. 2014, ApJ, 780, 127
\bibitem[\protect\citeauthoryear{}{}]{} Balbus, S. A., \& Hawley, J. F. 1991, ApJ, 376, 214
%\bibitem[\protect\citeauthoryear{}{}]{} Beckwith, K., Hawley, J. F., \& Krolik, J. H. 2008, ApJ, 678, 1180
\bibitem[\protect\citeauthoryear{}{}]{} Bu, D., Yuan, F., Wu, M., \& Cuadra, J. 2013, MNRAS, 434, 1692
\bibitem[\protect\citeauthoryear{}{}]{} Bu, D., Yuan, F., Gan, Z., \& Yang, X. 2016a, ApJ, 818, 83
\bibitem[\protect\citeauthoryear{}{}]{} Bu, D., Yuan, F., Gan, Z., \& Yang, X. 2016b, ApJ, 823, 90
\bibitem[\protect\citeauthoryear{}{}]{} Bu, D., \& Gan, Z. 2018, MNRAS, 474, 1206
\bibitem[\protect\citeauthoryear{}{}]{} Bu, D., Qiao, E., \& Yang, X. 2019, ApJ, 875, 147 (Paper I)
\bibitem[\protect\citeauthoryear{}{}]{} Burke, M. J., Gilfanov, M., \& Sunyaev, R. 2017, MNRAS, 466, 194
\bibitem[\protect\citeauthoryear{}{}]{} Burke, M. J., Gilfanov, M., \& Sunyaev, R. 2018, MNRAS, 474, 760
\bibitem[\protect\citeauthoryear{}{}]{} Campana, S., Brivio, F., Degenaar, N., et al. 2014, MNRAS, 441, 1984
\bibitem[\protect\citeauthoryear{}{}]{} Cheung, E., Bundy, K., Cappellari, M., et al. 2016, Nature, 533, 504
\bibitem[\protect\citeauthoryear{}{}]{} Clarke, D. 1996, ApJ, 457, 291
\bibitem[\protect\citeauthoryear{}{}]{} Crenshaw, D. M., \& Kraemer, S. B. 2012, ApJ, 753, 75
\bibitem[\protect\citeauthoryear{}{}]{} Degenaar, N., Wijnands, R., \& Miller, J. M. 2013a, ApJ, 767, L31
\bibitem[\protect\citeauthoryear{}{}]{} Degenaar, N., Miller, J. M., Wijnands, R., et al. 2013b, ApJ, 767, L37
%\bibitem[\protect\citeauthoryear{}{}]{} De Villiers, J. P., Hawley, J. F., \& Krolik, J. H. 2003, ApJ, 599, 1238
\bibitem[\protect\citeauthoryear{}{}]{} Done, C., Gierlinski, M., \& Kubota, A. 2007, A\&ARv, 15, 1
%\bibitem[\protect\citeauthoryear{}{}]{} Esin, A. A., Narayan, R., Ostriker, E., \& Yi, I. 1996, ApJ, 465, 312
\bibitem[\protect\citeauthoryear{}{}]{} Garcia, M. R., McClintock, J. E., Narayan, R., et al. 2001, ApJ, 553, L47
\bibitem[\protect\citeauthoryear{}{}]{} Gilfanov, M. R., \& Sunyaev, R. 2014, Phys-Usp., 57, 377
\bibitem[\protect\citeauthoryear{}{}]{} Gladstone, J., Done, C., \& Gierli\'{n}ski, M. 2007, MNRAS, 378, 13
\bibitem[\protect\citeauthoryear{}{}]{} Gu, W. 2015, ApJ, 799, 71
\bibitem[\protect\citeauthoryear{}{}]{} Hameury, J. M., Barret, D., Lasota, J. P., et al. 2003, A\&A 399, 631
%\bibitem[\protect\citeauthoryear{}{}]{} Hawley, J. F., Balbus, S. A., \& Stone, J. M. 2001, ApJ, 554, L49
%\bibitem[\protect\citeauthoryear{}{}]{} Hayes, J. C., Norman, M. L., Fiedler, R. A., et al. 2006, ApJ, 165, 188
\bibitem[\protect\citeauthoryear{}{}]{} Hern\'andez Santisteban, J. V., C\'uneo, V., Degenaar, N. et al. 2019, MNRAS, 488, 4596
\bibitem[\protect\citeauthoryear{}{}]{} Homan, J., Neilsen, J., Allen, J. L., et al. 2016, ApJ, 830, L5
\bibitem[\protect\citeauthoryear{}{}]{} Igumenshchev, I. V., \& Abramowicz, M. A. 2000, ApJ, 130, 463
\bibitem[\protect\citeauthoryear{}{}]{} Inogamov, N. A., \& Sunyaev, R. A. 1999, Astron. Lett., 25, 269
\bibitem[\protect\citeauthoryear{}{}]{} Jonker, P. G., Galloway, D. K., McClintock, J. E., et al. 2004, MNRAS, 354, 666
\bibitem[\protect\citeauthoryear{}{}]{} Kubota, A., \& Done, C. 2004, MNRAS, 353, 980
\bibitem[\protect\citeauthoryear{}{}]{} Lasota, J. P. 2000, A\&A, 360, 575
%\bibitem[\protect\citeauthoryear{}{}]{} Li, J., Ostriker, J., \& Sunyaev, R. 2013, ApJ, 767, 105
\bibitem[\protect\citeauthoryear{}{}]{} Ma, R., Roberts, S. R., Li, Y., \& Wang, Q. D. 2019, MNRAS, 483, 5614
%\bibitem[\protect\citeauthoryear{}{}]{} Maccarone, T. J. 2003, A\&A, 409, 697
\bibitem[\protect\citeauthoryear{}{}]{} Maccarone, T. J., \& Coppi, P. S. 2003, MNRAS, 338, 189
%\bibitem[\protect\citeauthoryear{}{}]{} Machida, M., Matsumoto, R., \& Mineshige, S. 2001, PASJ, 53, L1
%\bibitem[\protect\citeauthoryear{}{}]{} Marshall, M. D., Avara, M. J., \& McKinney, J. C. 2018, MNRAS, 478, 1837
\bibitem[\protect\citeauthoryear{}{}]{} McClintock, J. E., Narayan, R., \& Rybicki, G. B. 2004, ApJ, 615, 402
%\bibitem[\protect\citeauthoryear{}{}]{} McKinney, J. C., Tchekhovskoy, A., \& Blandford, R. D. 2012, MNRAS, 423, 3083
%\bibitem[\protect\citeauthoryear{}{}]{} McKinney, J. C., Dai, L., \& Avara, M. J. 2015, MNRAS, 454, L6
\bibitem[\protect\citeauthoryear{}{}]{} Medvedev, M. V., \& Narayan, R. 2001, ApJ, 554, 1255
\bibitem[\protect\citeauthoryear{}{}]{} Medvedev, M. V. 2004, ApJ, 613, 506
\bibitem[\protect\citeauthoryear{}{}]{} Menou, K., Esin, A. A., Narayan, R., et al. 1999, ApJ, 520, 276
\bibitem[\protect\citeauthoryear{}{}]{} Meyer-Hofmeister, E., Liu, B., \& Meyer, F. 2005, A\&A, 432, 181
%\bibitem[\protect\citeauthoryear{}{}]{} Morales Teixeira, D., Avara, M. J., \& McKinney, J. C. 2018, MNRAS, 480, 3547
\bibitem[\protect\citeauthoryear{}{}]{} Narayan, R., \& Popham, R. 1993, Nature, 362, 820
\bibitem[\protect\citeauthoryear{}{}]{} Narayan, R., \& Yi, I. 1994, ApJ, 428, L13
\bibitem[\protect\citeauthoryear{}{}]{} Narayan, R., \& Yi, I. 1995, ApJ, 452, 710 (NY95)
\bibitem[\protect\citeauthoryear{}{}]{} Narayan, R., \& McClintock, J. E. 2008, New Astron. Rev., 51, 733
\bibitem[\protect\citeauthoryear{}{}]{} Narayan, R., Sadowski, A., Penna, R. F., \& Kulkarni, A. K. 2012, MNRAS, 426, 3241
\bibitem[\protect\citeauthoryear{}{}]{} Nowak, M. A., Wilms, J., \& Dove, J. B. 2002, MNRAS, 332, 856
\bibitem[\protect\citeauthoryear{}{}]{} Park, J., Hada, K., Kino, M., et al. 2019, ApJ, 871, 257
%\bibitem[\protect\citeauthoryear{}{}]{} Pang, B., Pen, U. L., Matzner, C. D., et al. 2011, MNRAS, 415, 1228
\bibitem[\protect\citeauthoryear{}{}]{} Papaloizou, J. C. B., \& Pringle, J. E. 1984, MNRAS, 208, 721
%\bibitem[\protect\citeauthoryear{}{}]{} Pen, U. L., Matzener, C. D., \& Wong, S. 2003, ApJ, 596, L207
\bibitem[\protect\citeauthoryear{}{}]{} Popham, R., \& Narayan, R. 1992, ApJ, 394, 255
\bibitem[\protect\citeauthoryear{}{}]{} Popham, R., \& Sunyaev, R. 2001, ApJ, 547, 355
\bibitem[\protect\citeauthoryear{}{}]{} Pringle, J. E., Rees, M. J., \& Pacholczyk, A. G. 1973, A\&A, 29, 179
\bibitem[\protect\citeauthoryear{}{}]{} Qiao, E., \& Liu, B. 2009, PASJ, 61, 403
\bibitem[\protect\citeauthoryear{}{}]{} Qiao, E., \& Liu, B. 2013, ApJ, 764, 2
\bibitem[\protect\citeauthoryear{}{}]{} Qiao, E., \& Liu, B. 2018, MNRAS, 481, 938
\bibitem[\protect\citeauthoryear{}{}]{} Rodriguez, J., Corbel, S., \& Tomsick, J. A. 2003, ApJ, 595, 1032
%\bibitem[\protect\citeauthoryear{}{}]{} Ryan, B. R., Ressler, S. M., Dolence J. C., et al. 2017, ApJ, 844, L24
\bibitem[\protect\citeauthoryear{}{}]{} Sazonov, S. Y., Ostriker, J. P., Ciotti, L., \& Sunyaev, R. A. 2005, MNRAS, 358, 168
\bibitem[\protect\citeauthoryear{}{}]{} Sadowski, A., Naryan, R., Penna, R., \& Zhu, Y. 2013, MNRAS, 436, 3856
%\bibitem[\protect\citeauthoryear{}{}]{} Sadowski, A., Wielgus, M., Narayan, R., et al. 2017, MNRAS, 466, 705
\bibitem[\protect\citeauthoryear{}{}]{} Shakura, N. I., \& Sunyaev, R. A. 1973, A\&A, 24, 337
\bibitem[\protect\citeauthoryear{}{}]{} Stone, J. M., Pringle, J. E., \& Begelman, M. C. 1999, MNRAS, 310, 1002
%\bibitem[\protect\citeauthoryear{}{}]{} Stone, J. M., \& Pringle, J. E. 2011, MNRAS, 322, 461
\bibitem[\protect\citeauthoryear{}{}]{} Sunyaev, R. A., \& Titarchuk, L. 1989, in Hunt J., Battrick B. eds, Two Topics in X-Ray Astronomy, Volume 1: X Ray Binaries. Volume 2: AGN and the X Ray Background, ESA SP-296
\bibitem[\protect\citeauthoryear{}{}]{} Sunyaev, R. A., et al. 1991, Sov. Astron. Lett., 17, 409
\bibitem[\protect\citeauthoryear{}{}]{} Tchekhovskoy, A., Narayan, R., \& McKinney, J. C. 2011, MNRAS, 418, L79
\bibitem[\protect\citeauthoryear{}{}]{} van den Eijnden, J., et al. 2018, MNRAS, 475, 2027
\bibitem[\protect\citeauthoryear{}{}]{} Vaughan, B. A., van der Klis, M., Wood, K. S., et al. 1994, ApJ, 435, 362
\bibitem[\protect\citeauthoryear{}{}]{} Wang, Q. D., Nowak, M. A., Markoff, S. B., et al., 2013, Sci, 341, 981
%\bibitem[\protect\citeauthoryear{}{}]{} Wijnands, R., Degenaar, N., Armas Padilla, M. et al. 2015, MNRAS, 454, 1371
%\bibitem[\protect\citeauthoryear{}{}]{} Xie, F., \& Yuan, F. 2012, MNRAS, 427, 1580
\bibitem[\protect\citeauthoryear{}{}]{} Yuan, F. 2001, MNRAS, 324, 119
\bibitem[\protect\citeauthoryear{}{}]{} Yuan, F., \& Zdziarski, A. 2004, MNRAS, 354, 953
\bibitem[\protect\citeauthoryear{}{}]{} Yuan, F., Zdziarski, A., Xue, Y., \& Wu, X. 2007, ApJ, 659, 541
\bibitem[\protect\citeauthoryear{}{}]{} Yuan, F., \& Narayan, R. 2014, ARA\&A, 52, 529
\bibitem[\protect\citeauthoryear{}{}]{} Yuan, F., Bu, D., \& Wu, M. 2012, ApJ, 761, 130
\bibitem[\protect\citeauthoryear{}{}]{} Yuan, F., Gan, Z., Narayan, R., et al. 2015, ApJ, 804, 101
\bibitem[\protect\citeauthoryear{}{}]{} Zampieri, L., Turolla, R., Zane, S., et al. 1995, ApJ, 439, 849
\bibitem[\protect\citeauthoryear{}{}]{} Zhang, Z., Sakurai, S., Makishima, K., et al. 2016, ApJ, 823, 131

\end{thebibliography}
\end{document}